\documentstyle[12pt,epsf]{article}

\def\R{{\cal R}}

\def\beq{\begin{equation}}
\def\eeq{\end{equation}}
\def\ba{\begin{eqnarray}}
\def\ea{\end{eqnarray}}

\def\yt{\widetilde y}

\def\Q2t{{\widetilde Q}^2}

\def\k0t{\widetilde{k}_0}

\def\magk{|{\bf k}|}
\def\magq{|{\bf q}|}
\def\lsim{\mathrel{\rlap{\lower4pt\hbox{\hskip1pt$\sim$}}
		   \raise1pt\hbox{$<$}}}
\def\gsim{\mathrel{\rlap{\lower4pt\hbox{\hskip1pt$\sim$}}
		   \raise1pt\hbox{$>>$}}}

\def\ci{\cite}

\begin{document}
\begin{center}
\large
{\bf Many-body theory interpretation of deep inelastic scattering} \\
\normalsize
\vspace*{5mm}
 O. Benhar$^{1}$,  V.R. Pandharipande$^2$, I. Sick$^3$ \\
\vspace*{3mm}
$^1$ INFN, Sezione Roma 1, I-00185 Rome, Italy \\
$^2$ Department of Physics, University of Illinois, Urbana, IL 61801, USA\\
$^3$ Dept. f\"ur Physik und Astronomie,
Universit\"at Basel, CH-4056 Basel, Switzerland \\

\date{\today}
\end{center}
\normalsize

ABSTRACT: We analyze data on deep inelastic scattering of electrons from the proton
using ideas from standard many-body theory involving {\em bound} constituents
subject to {\em interactions}. This leads us to expect, at large three-momentum
transfer ${\bf{q}}$, scaling in terms of the variable  $\tilde{y}=\nu-{\bf |q|}$.
The response at constant ${\bf |q|}$ scales well in this variable. Interaction
effects are manifestly displayed in this approach. They are illustrated in two
examples. \\[5mm]
The cross section for deep inelastic  scattering (DIS) of electrons  on 
unpolarized 
protons (and, {\em mutatis mutandis}, neutrons) is usually expressed as 
\cite{Halzen84}: 
\beq
\frac{d^2\sigma}{d\Omega d\nu} = \sigma_M \left[ 2 W_1 (|{\bf q}|, \nu) 
tan^2 \frac{\theta}{2} + W_2(|{\bf q}|, \nu) \right]\ , 
\label{eq:cs}
\eeq
where $\sigma_M$ is the  Mott  cross section, $\theta$ the scattering 
angle and $\nu$ the energy transfer. In this brief note we discuss only 
$W_1$. 

$W_1$ is generally considered as a function of $Q^2 = |{\bf q}|^2 - \nu^2$ 
and Bjorken $x = Q^2/2m \nu$ for scattering by protons initially 
at rest in laboratory frame.  The  advantages of using the Lorentz scalar 
variables $Q^2$ and $x$ are discussed in standard texts \cite{Halzen84}. 
The data show that $W_1(Q^2,x)$ obeys Bjorken scaling at large values of 
$Q^2$; it depends primarily on $x$. 
The weak dependence of $W_1$ on $Q^2$ is well understood via the 
perturbative QCD theory developed by Gribov, Lipatov,  
Altarelli and Parisi (GLAP) \cite{Ellis96}. The DIS data are usually
interpreted by going to the infinite momentum frame where $x$ is identified
as the fraction  of the momentum  carried by the quark responsible 
for the deep inelastic scattering. This interpretation has been very 
helpful in understanding DIS and in interpreting 
high energy reactions.  

\begin{figure}[htb]
\centerline{\mbox{\epsfysize=7cm\epsffile{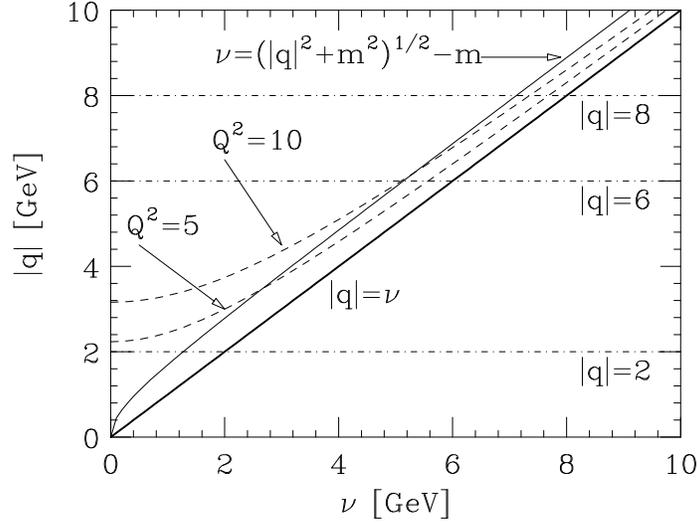}}}
\begin{center} \parbox{11cm}{\caption{ 
\label{qnuplane} 
$|{\bf q}|, \nu $ plane. The thick solid line separates space- and timelike 
region. The solid line corresponds to elastic e-p scattering, the dashed lines
to constant four-momentum transfer. 
 } }  
\end{center}
\end{figure}

In many-body theory it is natural to study the response of a system 
in its rest frame at fixed values of the momentum transfer ${\bf q}$ 
as a function of the energy transfer $\nu$.   The response to a scalar 
probe, for example, is viewed as
the distribution of the strength of the state $\sum_i e^{i {\bf q} \cdot 
{\bf r_i}}|0\rangle$, created by the probe, 
among the eigenstates of the system belonging
to momentum ${\bf q}$.  Fig.\ref{qnuplane} 
shows the various domains of the response of a proton in the $|{\bf q}|$, 
$\nu$ plane.  The thick line $|{\bf q}|=\nu$ separates the spacelike response 
above the line, and the time like below the line.  The thin line shows the 
$e-p$ elastic scattering kinematical condition: 
\beq
\nu_{el} = \sqrt{(|{\bf q}|^2 + m^2)} - m.  
\label{eq:nuel}
\eeq
At large values of $|{\bf q}|$ the $\nu_{el}=|{\bf q}|-m$ up to terms of 
order $m^2/2|{\bf q}|$.  There can not be any response above the line 
$\nu_{el}(|{\bf q}|)$ since none of the target states can have 
energy less than $\sqrt{(|{\bf q}|^2 + m^2)}$.  The dashed lines show parabolae: 
$|{\bf q}|^2 - \nu^2 = Q^2$ for $Q^2 = 5$ and 10 GeV$^2$.  They intersect 
the $\nu_{el}(|{\bf q}|)$ curve at $x=1$ and approach the $\nu = |{\bf q}|$ 
line at $\nu \rightarrow \infty$ or equivalently as $x \rightarrow 0$.
In most of the literature the authors have considered the 
variation of $W_1$ along these parabolae which do not enter the timelike 
region.  Here we study the variation of $W_1$ along the dash-dot lines, which 
have constant $|{\bf q}|$ and enter the timelike region. 

\begin{figure}[htb]
\centerline{\mbox{\epsfysize=7cm \epsffile{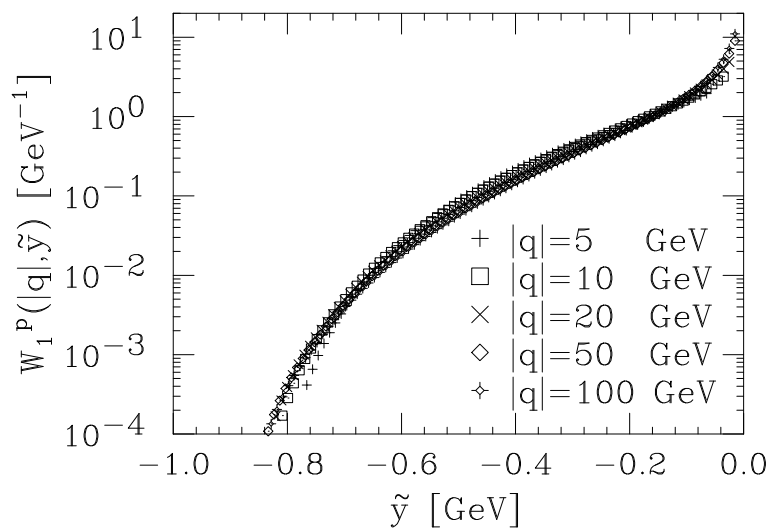}}}
\begin{center} \parbox{11cm} {\caption{ 
\label{w1pqnu} 
W$_1$ of the proton as a function of the scaling variable $\tilde{y} = \nu - |\bf{q}|$
for a range of momentum transfers.
} } 
\end{center}
\end{figure}

In fig.\ref{w1pqnu} we show the proton $W^p_1({\bf q},\nu )$, obtained from 
the MRS(A) fit of ref.\cite{Martin95} to $e$-$p$ 
and other cross sections, at several 
values of $|{\bf q}|$ as a function of $\nu - |{\bf q}| \equiv \yt$. 
The data show that at large values of 
${\bf q}$ the $W^p_1({\bf q},\nu )$ depends primarily on $ \yt $. 
This scaling has a simple interpretation in the many-body theory, 
related to the well known $y$-scaling \cite{Day90}, as will be discussed below.   
Since $\nu - |{\bf q}| = -m \xi $, where $\xi$ is the Nachtmann 
\cite{Nachtmann73,Jaffe85} 
scaling variable, it is closely related to Bjorken scaling as well.  
The Nachtmann variable, which results from operator product expansion studies, 
coincides with $x$ at large $Q^2$ and is generally used to extend the 
applicability of Bjorken scaling to lower $Q^2$. The small scaling 
violations seen in fig.\ref{w1pqnu}  originate from the gluonic radiative 
corrections as in the standard approach based on GLAP evolution equations
\cite{Ellis96}.  Contrary to the case of $x$-scaling, in
$ \yt $  scaling  $Q^2$ has a large variation, from zero at $|{\bf q}| = 
\nu $ ($\yt=0) $ to $\sim 2m|{\bf q}|$ at $\nu = |{\bf q}|-m$ ($\yt=m$), 
which does not appear to spoil the quality of scaling.  

Many-body theory views the deep inelastic response as follows. The 
probe creates states 
$|i({\bf q}+{\bf k});\R(-{\bf k}) \rangle $  by hitting a bound constituent   
$i$ of the system with initial momentum ${\bf k}$ and the residual system in the 
state $\R$ with momentum $-{\bf k}$.  In the plane wave impulse 
approximation (PWIA) the final state interaction (FSI) between the struck constituent
$i$ and the residual system $\R$ are neglected.  In this approximation the 
energy of the state $|i({\bf q}+{\bf k});\R(-{\bf k}) \rangle $ is: 
\beq
E(i({\bf q}+{\bf k});\R(-{\bf k})) = |{\bf q}| + k_{\parallel} + 
E(\R(-{\bf k})) +\  terms\  of \  order \ \frac{1}{|{\bf q}|} \ ,
\label{eq:eirk}
\eeq
where $k_{\parallel}$ is the projection of ${\bf k}$ in the direction of 
${\bf q}$.  At large $\magq$ the terms of order $1/|{\bf q}|$ can be 
neglected, and the response due to the excitation of the state 
$|i({\bf q}+{\bf k});\R(-{\bf k}) \rangle $ occurs at energy transfer: 
\beq
\nu = E(i({\bf k}+{\bf q});\R(-{\bf k}))-m = |{\bf q}| + {\bf k}_\parallel + 
E(\R(-{\bf k}))-m \ .
\label{eq:nuirk}
\eeq
Since the $\nu - |{\bf q}|$ is independent of 
${\bf q}$ at large $\magq$, the response depends only on $ \yt $ in 
the PWIA; therefore it scales. 

The fact that the observed response of the proton, as seen in 
fig.\ref{w1pqnu}, scales with $ \yt $ does not necessarily 
imply that PWIA is valid.  In general the effects of the FSI on the 
response may not be negligible. Treatments of the FSI for very different cases
of inclusive scattering of a probe from a composite system
\cite{Silver87,Benhar99,Carraro91} have shown that the main effect of FSI
results in a folding of the PWIA response:
\beq
W_1(\magq,\nu) = \int d \nu^{\prime} W_{1,PWIA} (\magq,\nu^{\prime}) 
f(\magq,\nu,\nu^{\prime})\ .
\label{eq:fold}
\eeq
The scaling of $W_1(\magq,\nu)$ with $ \yt $ can occur when 
the folding function representing the effect of the final state 
interactions becomes independent of $\magq$.  For example, in the 
Glauber approximation, the folding function for the quasi-free 
scattering of electrons by nuclei becomes independent of $\magq$ 
at $\magq > 2$ GeV, as has been recently discussed by Benhar
\cite{Benhar99}.  Weinstein and Negele \cite{Weinstein82} have shown 
that an analogous $y$-scaling in hard sphere Bose gas occurs even 
though the FSI effects are strong. 

By boosting the state $|\R(-{\bf k}),i({\bf k}) \rangle$ to large velocity,
ignoring interaction between $i$ and $\R$, it can be shown that 
\beq
\xi = \frac{|{\bf q}|-\nu}{m} = 1 - \frac{E(\R(-{\bf k}))+k_\parallel}{m} 
\eeq
has the usual meaning of the fraction of the momentum carried by the 
struck particle i.  

The complete response is obtained by summing over all the final states. 
Therefore:
\beq
W_1({\bf q},\nu)= \sum_i \int d^3k~de~ \sigma_1({\bf q},\nu,{\bf k},e)
P_i({\bf k},e) \delta(\yt - k_{\parallel} - e) \ ,
\label{eq:pwia1}
\eeq
where the spectral function $P_i({\bf k},e)$ is given by:
\beq
P_i({\bf k},e)= \sum_{\R} |\langle \R(-{\bf k}),i({\bf k})|0 \rangle |^2 
\delta(m - E(\R(-{\bf k})) - e)\ .
\label{eq:spf}
\eeq
The ${\bf k},e$ can be considered as the initial momentum and energy of 
the struck particle, which is treated as {\em bound}, and therefore not 
on the mass shell.  The $\sigma_1$ is the transverse cross-section for lepton 
scattering by a bound spin 1/2 point particle $i$, divided by $\sigma_M$. 
In the physical spacelike $(\yt < 0)$ region, $\sigma_1 = q_i^2$ in the 
large ${\bf q}$ limit, neglecting the mass $m_i$ of $i$.  Here $q_i$ is the 
charge of $i$.  This gives:
\beq
W_1({\bf q},\nu)= W_1(\yt) = \sum_i q_i^2 \int d^3k~de~ 
P_i({\bf k},e) \delta(\yt - k_{\parallel} - e) \ .
\label{eq:pwia2}
\eeq
This Eq. provides the relation between the familiar $F_1(\xi)=mW_1(\xi)$ 
structure function and the $P_i({\bf k},e)$ spectral function.  The 
$\delta$-function implies that $\yt = k_{\parallel} + e$, closely 
resembling the scaling variable $y$ used in  quasi-elastic electron-nucleus 
scattering, where it is associated with the parallel momentum of the 
struck nucleon. 

This simple picture of the response will be modified by the color 
confining interactions.  The mass of the nucleon contains confinement 
interaction contribution, while it is omitted in the energy of the struck quark,
$\magq + k_{\parallel}$.  It therefore must be included in the energy $E(\R)$ 
of the residual system.  We expect that the confinement energy does not 
change significantly in the time duration of the DIS, and its main 
influence is via the wave functions $|0\rangle$ and $|\R\rangle$.  
However, it could also contribute to the FSI folding function
(Eq.(\ref{eq:fold})). 

An interesting feature of fig.\ref{w1pqnu} concerns the width 
of the response, which amounts to only few hundred MeV independent of
the value of $\magq$.  This implies that deep 
inelastic scattering has an intrinsic energy scale of few hundred MeV. 
The main part of the transferred energy, of order $\magq$, goes into the 
kinetic energy of the struck constituent, and does not play any interesting 
role in the dynamics of the target system.  Therefore changes in the 
energy $E(\R(-{\bf k}))$ of the residual system of order 100 MeV have 
significant effect on the $W_1(\yt )$, or equivalently on 
$F_1( \xi )$.  In the following we discuss 
two observed effects of $E(\R(-{\bf k}))$ on the response. 

\begin{figure}[htb]
\centerline{\mbox{\epsfysize=7cm\epsffile{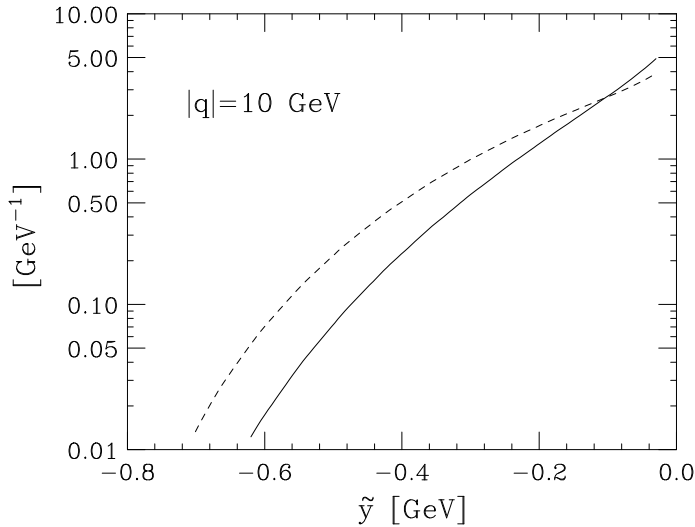}}}
\begin{center} \parbox{11cm}{\caption{ 
\label{ud} Valence quark responses $\chi_0$ (dashed) and $\chi_1$ (solid) for 
the proton. 
} }  
\end{center}
\end{figure}

The first, studied by Close and Thomas \cite{Close88}, concerns the difference 
between the responses due to valence u and d quarks in the proton.  Let 
$V_u(\yt )$ and $V_d(\yt )$ be the contributions of  
valence u and d quarks to the $W^p_1(\yt )$.  When the 
lepton strikes the valence d quark, the remaining two valence u quarks are 
left in the residual state $\R_1$ with spin 1.  In contrast, when a valence 
u quark is struck, the residual u-d pair is in states $\R_0$ with spin 0 
with probability 0.75, 
and $\R_1$ with probability 0.25.  Therefore $ \chi_1(\yt) \equiv 9 V_d(\yt)$ 
is the response due to final states $\R_1$ (normalized to unit particle charge), 
while $\chi_0(\yt) \equiv 1.5 (V_u(\yt)-2 V_d(\yt)$ is that for $\R_0$. 
The
\beq
E(\R_1(-{\bf k})) - E(\R_0(-{\bf k})) \sim \frac{2}{3}(m_{\Delta}-m) \ , 
\eeq
in perturbation theory, 
and therefore we expect $\chi_1(\yt)$ to be shifted by $\sim 0.2$
GeV from $\chi_0(\yt)$.  Fig.\ref{ud} 
shows that these responses obtained from the MRS(A) parton distributions at 
$\magq = 10$ GeV are indeed shifted by $\sim 0.1$ GeV from each other, 
at $\yt < -0.2$. In the PWIA, this shift should be independent of $\yt$, provided 
the color magnetic interaction can be treated perturbatively. The fact that the
shift is only $\sim$ 0.1 GeV indicates that it has nonperturbative 
contributions. Differences in FSI can also have an influence.

The second example concerns the modification of the 
deep inelastic response by nuclear 
effects \cite{Arneodo94} as first observed by the EMC collaboration \cite{Aubert83a}.
The EMC ratio $R_A(x)$ of the cross section per nucleon, 
for nucleus with mass number $A$ to that for the deuteron, 
does not show any $Q^2$ dependence within the experimental errors. 
The observed $R_A(x)$ has been extrapolated using Local Density Approximation 
to obtain the ratio $R_{NM}(x)$ for uniform nuclear matter
\cite{Sick92b}.
\begin{figure}[htb]
\centerline{\mbox{\epsfysize=7cm\epsffile{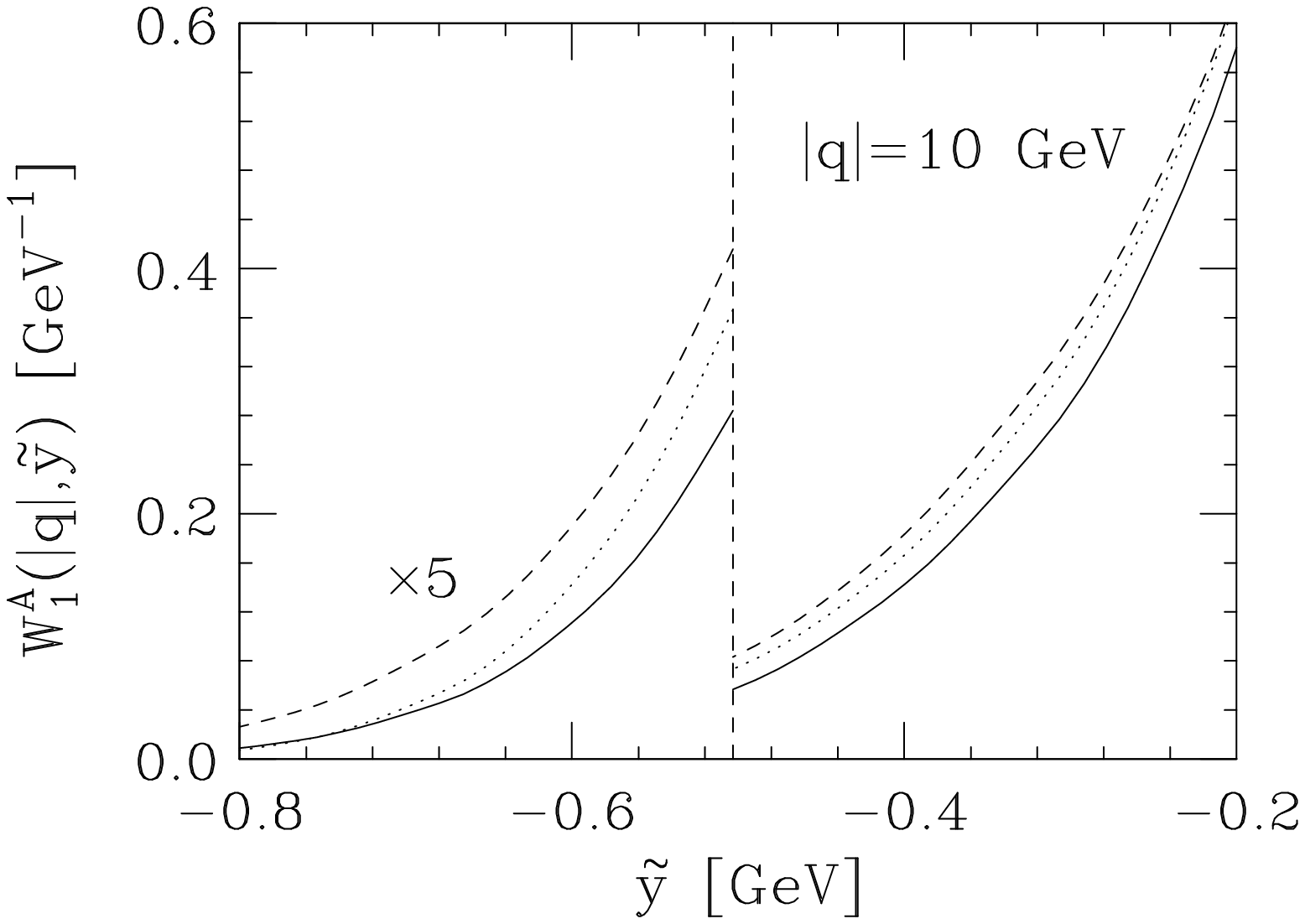}}}
\begin{center} \parbox{11cm}{\caption{ 
\label{w1a} 
Nuclear matter structure function (solid) compared to deuteron structure 
function (dotted) and structure function calculated for noninteracting, moving 
nucleons (dashed)
} }  
\end{center}
\end{figure}
In fig.\ref{w1a} we show the $W_1^d(\yt)$ for the deuteron 
calculated from the MRS(A) fits, and the $W_1^{NM}(\yt)$ 
at $\magq = 10$ GeV.  As we see from
this figure, the nuclear matter response is quite similar to that of the 
deuteron.  It is a bit broader,  due to the Fermi motion of 
nucleons in matter, but mainly  it is shifted towards higher $\nu$ due to nuclear binding.  
Fig.\ref{w1a} also shows the response of noninteracting nucleons 
distributed according to the momentum distribution of nucleons in 
nuclear matter, also calculated using realistic 
interactions \cite{Fantoni84}.  The observed response is shifted relative the 
Fermi motion broadened response by $\sim$ 40--60  MeV, which is 
comparable to the average nucleon removal energy of $\sim$ 62 MeV 
in nuclear matter \cite{Benhar89}. For these reasons, the conventional 
nuclear physics approach 
is quite successful in describing the EMC ratio for nuclear matter at 
$x > 0.4$ \cite{Benhar97}. 

In the parton model the struck particle is assumed to be on mass-shell before and
after the interaction with the electron \cite{Ellis83}. In this case 
\beq
\nu = \sqrt{m_i^2+({\bf k}+{\bf q})^2}-\sqrt{m_i^2+{\bf k}^2} \leq \magq\ ,
\label{eq:nuparton}
\eeq
and all of the response is at negative $\yt$, in the spacelike region. The same is valid 
at the leading twist-two order of the operator product expansion \ci{Jaffe85}.

In many body theory, however, a timelike response occurs either due to initial
state interactions, which  can make $E(\R(-{\bf k}))$ 
large enough to give a positive right hand side of eq.(\ref{eq:nuirk}),
 or because of FSI.  The initial energy of the struck 
constituent is identified with $e = m - E(\R(-{\bf k}))$
 and not with the 
on shell energy $\sqrt{m_i^2 + \magk^2}$ used in eq.(\ref{eq:nuparton}). 
This timelike response contributes to various sum rules.
For  example, the Coulomb sum in quasi-elastic electron nucleus scattering 
is defined as the integral of the 
longitudinal response over both space and time like regions. 
The longitudinal response of deuterium 
has been calculated with realistic forces \cite{Arenhoevel89}; it 
extends into the timelike region, and that region has to be 
included to fulfill the Coulomb sum.  

In fact, the shifts in $W_1(\yt)$ illustrated in figs.\ref{ud} and \ref{w1a}
will move part of the response into the timelike region, barring FSI effects, 
and thus lead to a violation of sum rules involving $W_1(\yt<0)$.

In conclusion, we obtain new insights in the deep inelastic response
of nucleons by applying standard many-body theory and relate the scaling
function to the nucleon spectral function in the lab frame. 
The natural scaling variable of many body theory, $\yt$, equals $-m\xi$ of 
the conventional approach to DIS. While $\yt$ scaling is derived 
 assuming bound
constituents which are subject to initial and final state interaction, 
$\xi$- or $x$-scaling is
obtained assuming free constituents without FSI. The occurrence of scaling thus
cannot automatically be taken as evidence for scattering from free 
constituents. 

Acknowledgements: The authors thank A.W. Thomas and D. Beck for interesting
discussions. 
This work was supported by the US and Swiss National Science Foundations. 
\bibliography{sick}
\bibliographystyle{unsrt}

\end{document}